\documentclass[fleqn,usenatbib]{mnras}
\usepackage{newtxtext,newtxmath}

\usepackage[T1]{fontenc}
\usepackage{ae,aecompl}

\usepackage{graphicx}	
\usepackage{amsmath}	
\usepackage{amssymb}	

\title[35 Myr old filamentary relic resolved by Gaia]{Uncovering a 260 pc wide, 35 Myr old filamentary relic of star formation}
\author[Beccari, Boffin \& Jerabkova]{Giacomo Beccari,$^{1}$\thanks{E-mail: gbeccari@eso.org}
Henri M.J. Boffin,$^{1}$ and
Tereza Jerabkova,$^{1,2,3,4,5,6}$ 
\\
$^{1}$European Southern Observatory, Karl-Schwarzschild-Strasse 2, 85748 Garching bei M\"unchen\\
$^{2}$Helmholtz Institut f\"{u}r Strahlen und Kernphysik, Universit\"{a}t Bonn, Nussallee 14--16, 53115 Bonn, Germany\\
$^{3}$Astronomical Institute, Charles University in Prague, V Hole\v{s}ovi\v{c}k\'ach 2, CZ-180 00 Praha 8, Czech Republic\\
$^{4}$ Astronomical Institute, Czech Academy of Sciences, Fri\v{c}ova 298, 25165, Ond\v{r}ejov, Czech Republic \\
$^{5}$ Instituto de Astrof{\'i}sica de Canarias, E-38205 La Laguna, Tenerife, Spain\\
$^{6}$ GRANTECAN, Cuesta de San Jose s/n, 38712 Brena Baja, La Palma, Spain\\
}

\date{Accepted 2019 November 13. Received 2019 November 13; in original form 2019 July 04}

\pubyear{2019}

\begin{document}
\label{firstpage}
\pagerange{\pageref{firstpage}--\pageref{lastpage}}
\maketitle

\begin{abstract}
Several recent studies have shown that the Vela OB2 region hosts a complex constellation of sub-populations with ages in the range 10 to 50 Myr. Such populations might represent the best example of the outcome of clustered star formation in Giant Molecular clouds (GMC). We use Gaia DR2 data over an area of 40 deg radius around the open cluster Collinder 135 to extend the study of the stellar populations of the Vela OB2 region over an area of several hundreds of parsecs on sky. Detailed clustering algorithms combined with the exquisite astrometric quality of the GAIA catalogue allow us to detect a new cluster named BBJ~1 that shows the same age as NGC~2547 (30 to 35 Myr), but located at a distance of 260 pc from it. Deeper investigation of the region via clustering in 5D parameter space and in the colour-magnitude diagram allows us to detect a filamentary structure of stars that bridges the two clusters. Given the extent in space of such structure (260 pc) and the young age ($\sim$35~Myr), we exclude that such population originates by the same mechanism responsible to create tidal streams around older clusters. Even if we miss a complete picture of the 3D motion of the studied stellar structure because of the lack of accurate radial velocity measurements, we propose that such structure represent the detection of a 35 Myr old outcome of a mechanism of filamentary star formation in a GMC.  

\end{abstract}




\section{Introduction}
Since the seminal study by~\citet{lada03}, it is generally understood that the vast majority of stars are formed in embedded star clusters.  
In particular, several observations in the infrared and sub-mm using the satellite Herschel and the {\it Atacama Large Millimeter/submillimeter Array} (ALMA) observatory\citep{Andre2014,Hacar2018} seem to indicate that 
most stars form in sub-parsec high-density regions in molecular clouds (MCs). 

Such high mass density sub-parsec regions in MCs are made of a complex structure of intersecting molecular small-scale gas fibers and larger-scale filaments \citep{Hacar+17b,Hacar2018}. The density along individual star-forming filaments is high enough ($10^5$ particles/cm$^3$) to gravitationally fragment into low mass stars. However, the vast majority of stellar cores are found at intersections of filaments and they seem to exhibit primordial mass segregation 
\citep{pl18,Pavlik2019}. As an example, \citet[][]{Joncour2018} recently found that nearly half of the entire stellar population in Taurus is concentrated in 20 very dense, tiny and prolate regions called NESTs (for Nested Elementary STructures).
Moreover the high-mass stars and embedded star clusters appear to be exclusively linked to places of high mass surface density and multiple intersections of filaments and fibers 
\citep{Schneider2012,Hennemann2012, Hacar2018}. 

By zooming out from individual sub-parsec embedded clusters, or NESTs of star formation, we can see that gas and dust filament complexes can have lengths up to 100pc. For example, \citet[][]{gro18} has recently used the position and astrometric information of the young stellar objects (younger than 3Myr) in Orion A and available from Gaia-DR2 to study the 3D shape and orientation of the giant molecular cloud (GMC) Orion A. They find that the true extent of Orion A is 90 pc, more than twice bigger with respect to the projected 40 pc.

Moreover the sub-structure of molecular clouds seem to be such that filaments \& fibers can create large filament complexes, usually called bones, which can be linked with galaxy-wide instabilities induced by bar or spiral arms \citep[e.g.][]{Li2016,Mattern2018}. In the era of Gaia the following question thus emerges: what imprint does this star formation in seemingly large scale filaments leave in young stellar populations just emerging from their natal molecular clouds? 

Wide field optical to near-infrared surveys revealed that when a star forming 
region is observed and studied at a scale of hundreds of parsecs it shows that young stars aggregate in multiple sub-clusters \citep{gro18,ks18,ziv18}. Very recently, the unprecedented
astrometric precision reached through the data taken by the Gaia satellite~\citep{ga16} and available to the community trough the second data release Gaia DR2~\citep{ga18b}, made it possible to isolate and study the stellar population in clusters with unequaled precision~\citep{zari18}. 

Along this line, the Vela OB2 association has recently revealed to be one of the best test-bench to study the early
stages of clustered star formation (SF) at ages 10 to 35~Myr. Using a combination of Gaia DR2 astrometric information and accurate
photometry, \citet[][]{becc18} recently  studied a $10\times5$ deg region around the Wolf-Rayet Star $\gamma^2$ Vel, 
including the well known clusters Gamma Vel and NGC2547. They discovered an assembly of six clusters. Four of them 
formed coevally from the same molecular clouds 10 Myr ago, while NGC 2547 formed together with a newly discovered cluster 30 to 35 Myr ago. 
Later~\citet[][]{ca19a,ca19b} expanded the study on a wider area ($\sim30\times30$~deg) within the Gum nebula. They hypothesise that the set of 10 Myr clusters including Vela OB2 have originated from a energetic event like a supernova explosion, which could be the cause of the IRAS Vela Shell. In this paper we extend the analysis of the stellar population to a wider 
region with the scope of investigating further any sign of formation history of the stars in the region dynamically imprinted in the 10 to 30 Myr complex stellar population. 

\section{The data set}

We first retrieved from the Gaia DR2 archive\footnote{\url{https://gea.esac.esa.int/archive/}} a catalogue 
of objects in a region of 40 degree radius around the
center of the cluster Collinder 135. Since we were particularly interested into a detailed study of the population in the Vela OB2 region, we restricted the selection of the stars to the range of parallaxes $1.5<\varpi<3.5$~\citep[see][]{cg18}. Moreover, following the prescription in \citet[][]{an17}, we restrict the study to only the objects for which the uncertainty of the parallax is lower than 10\%.

Initially, we visually inspected the distribution of the stars on the celestial map in right ascension R.A. and declination Dec. As expected, we could recognize many aggregates and clumps of stars likely associated with well know star clusters. We used the catalogue of open clusters' members recently published by~\citet[][]{cg18} to identify the position of the know clusters in 3D space in the surveyed area. Surprisingly, a clump of stars roughly centered at RA=06:21:08 Dec=-16:10:00 ($l$=224.221369, $b$=-13.866467) escaped the selection done by~\citet[][]{cg18}. No star clusters are found in the close vicinity in the recent work from~\citet[][]{bica19}. The search for star cluster-like objects in a 1$^\circ$ radius around the position of the cluster using Simbad~\citep{we00} gives negative result. Moreover no cluster appears at this position in the WEBDA\footnote{\url{http://www.univie.ac.at/webda/}} data base.

\begin{figure*}
\centering
 \includegraphics[width=0.32\hsize]{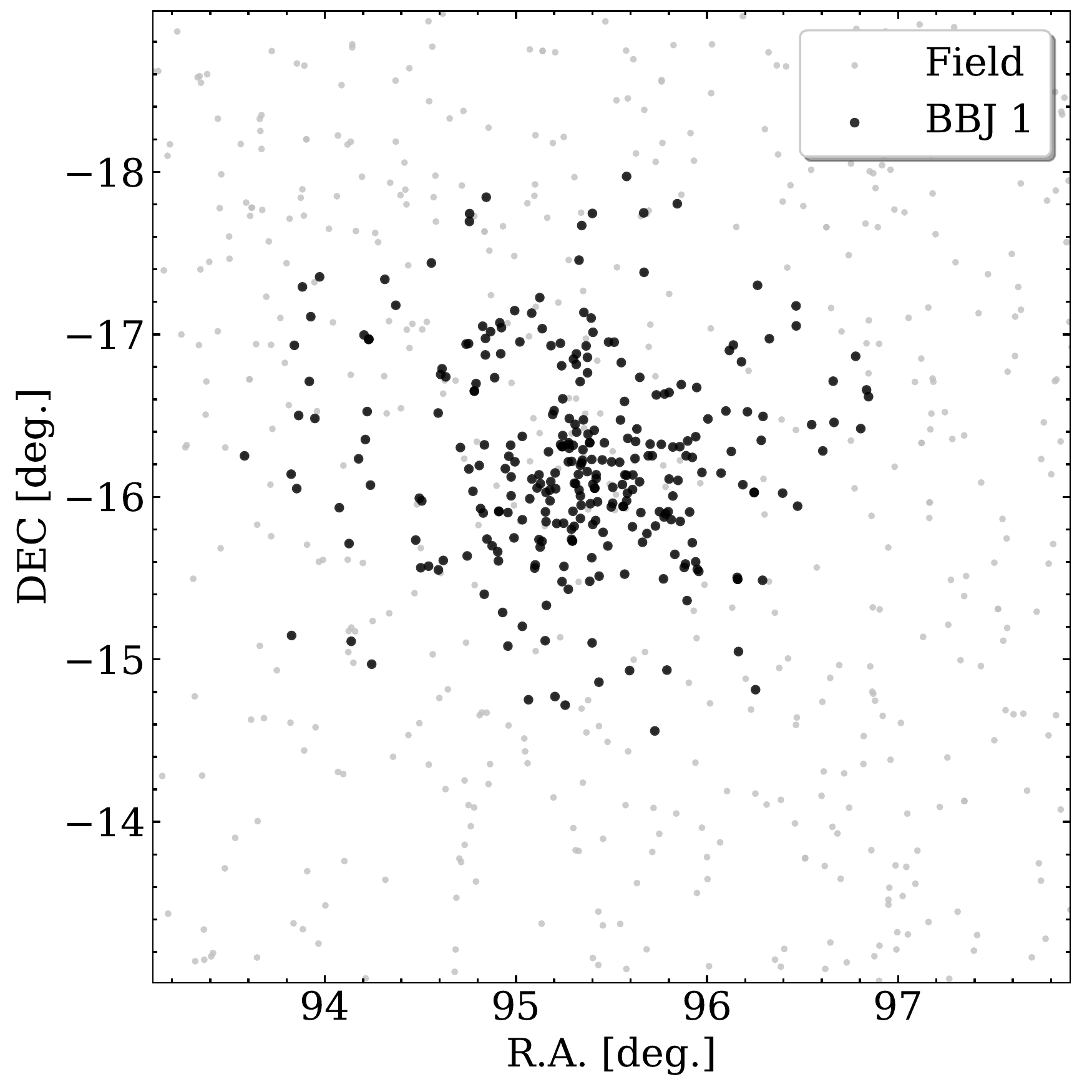}
 \includegraphics[width=0.32\hsize]{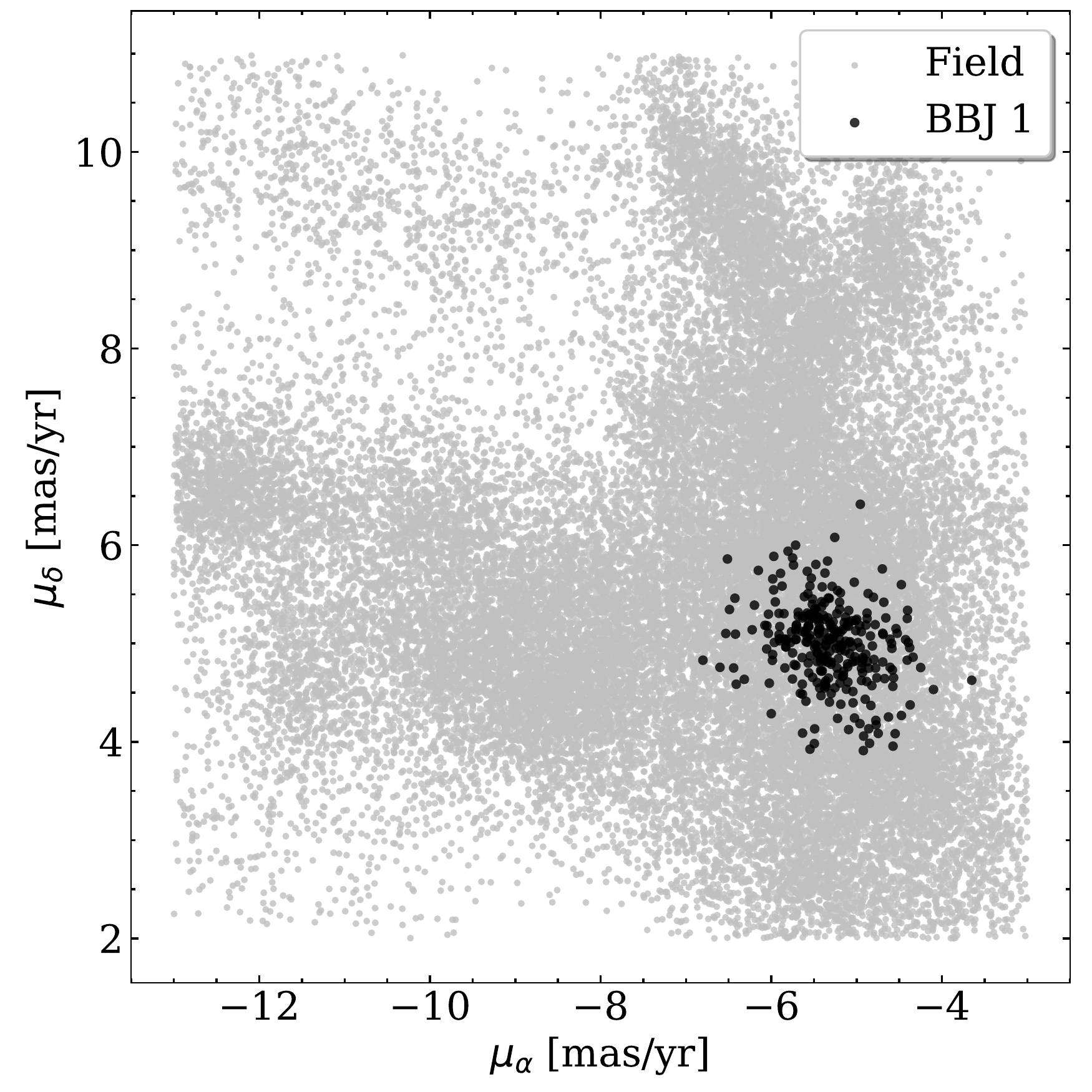}
 \includegraphics[width=0.32\hsize]{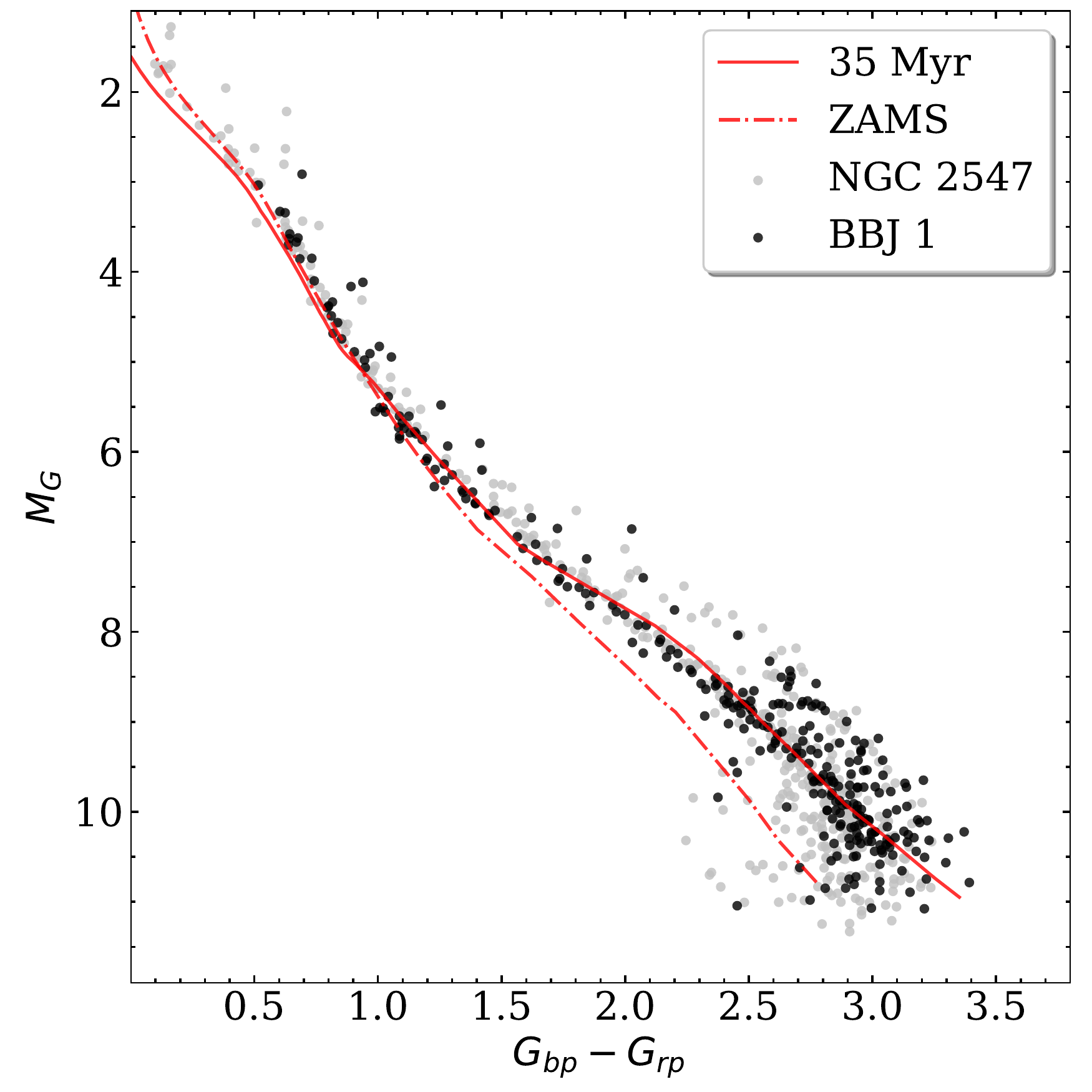}
  \caption{\textit{Left-panel}: Map of the stars in the region around the new cluster. Black dots indicate the stars identified as cluster members via the clustering algorithm while in grey we show the field stars \textit{Central-panel}: The distribution of the stars in the proper motion plane. The colour code is the same as on the left panel. \textit{Right-panel}: Comparison on the CMD of the stars belonging to BBJ 1 (black dots) and NGC 2547 (grey dots). The two clusters are clearly coeval.}
  \label{fig:bec1}
\end{figure*}

\section{A new star cluster: BBJ 1}

We used the DBSCAN data clustering algorithm \citep{Ester96adensity-based}
implemented as part of the python scikit-learn project\footnote{\url{http://scikit-learn.org/stable/modules/generated/sklearn.cluster.DBSCAN.html}} \citep{scikit-learn} in order to identify the presence of a star cluster at the position indicated above. DBSCAN requires the user to control two input parameters, namely a search radius (eps) and the minimum number of points required to form a cluster (min\_samples). In short the algorithm begins with an arbitrary starting point. Then, the point's eps-neighborhood is retrieved and if it contains a sufficient number  of  points,  a  cluster  forms. The  DBSCAN  algorithm  does  not  require  one  to  specify  the number  of  clusters,  and  it  can find  arbitrarily  shaped  clusters  in  parameter  space~\citep[][]{Gao_2014}.

In order to take full advantage of the astrometric information provided by the Gaia DR2 catalogue, we performed the clustering algorithm in 5D space, hence running the search simultaneously in $\alpha$, $\delta$, $\varpi$, $\mu_\alpha$ and $\mu_\delta$. On top of the selection of stars in $\varpi$ mentioned above, we further restrict the clustering search to an area in Galactic coordinates $222<l<225$ and $-15<b<-13$ and  $-8<\mu_\alpha<-3$ and $3<\mu_\delta<8$. We run DBSCAN using eps=0.550 and min\_samples=10.

The clustering method allowed us to clearly identify the new cluster. The population of stars identified by the DBSCAN algorithm as bona-fide cluster's members are shown as black solid points on the left panel of Fig.~\ref{fig:bec1}. 
As shown on the central panel of Fig.~\ref{fig:bec1} the stars do not only cluster in the coordinate space, but in proper motion around $\mu_\alpha=-5$ and $\mu_\delta=5$ mas/yr. This fact confirms that we have identified an assembly of stars clustered in space and dynamically correlated.
We show in the right panel of Fig.~\ref{fig:bec1} the $M_{G}$,$G_{bp}-G_{rp}$ colour-magnitude diagram (CMD) of the stars belonging to the newly discovered population (black dots) over-plotted to the stars belonging to the $\sim$30~Myr cluster NGC 2547~\citep[][]{bossini19,cg18}. The comparison confirms that the stars selected in 5D space using the DBSCAN clustering algorithm in fact do belong to a $\sim$30~Myr single stellar population. We call the new cluster BBJ 1. The distribution of parallax of the stars belonging to BBJ 1 peaks at $\varpi\sim2.67\pm0.067$ i.e. $\sim$374~pc. This fact places the cluster at the same distance of the several clusters and association somehow related to the Vela OB2 complex, including NGC~2547~\citep[384pc][]{ca19b}.

\section{Filamentary structures in the Vela OB2 complex} 
\label{sec:clustering}
\begin{figure*}
\centering
 \includegraphics[width=0.9\hsize]{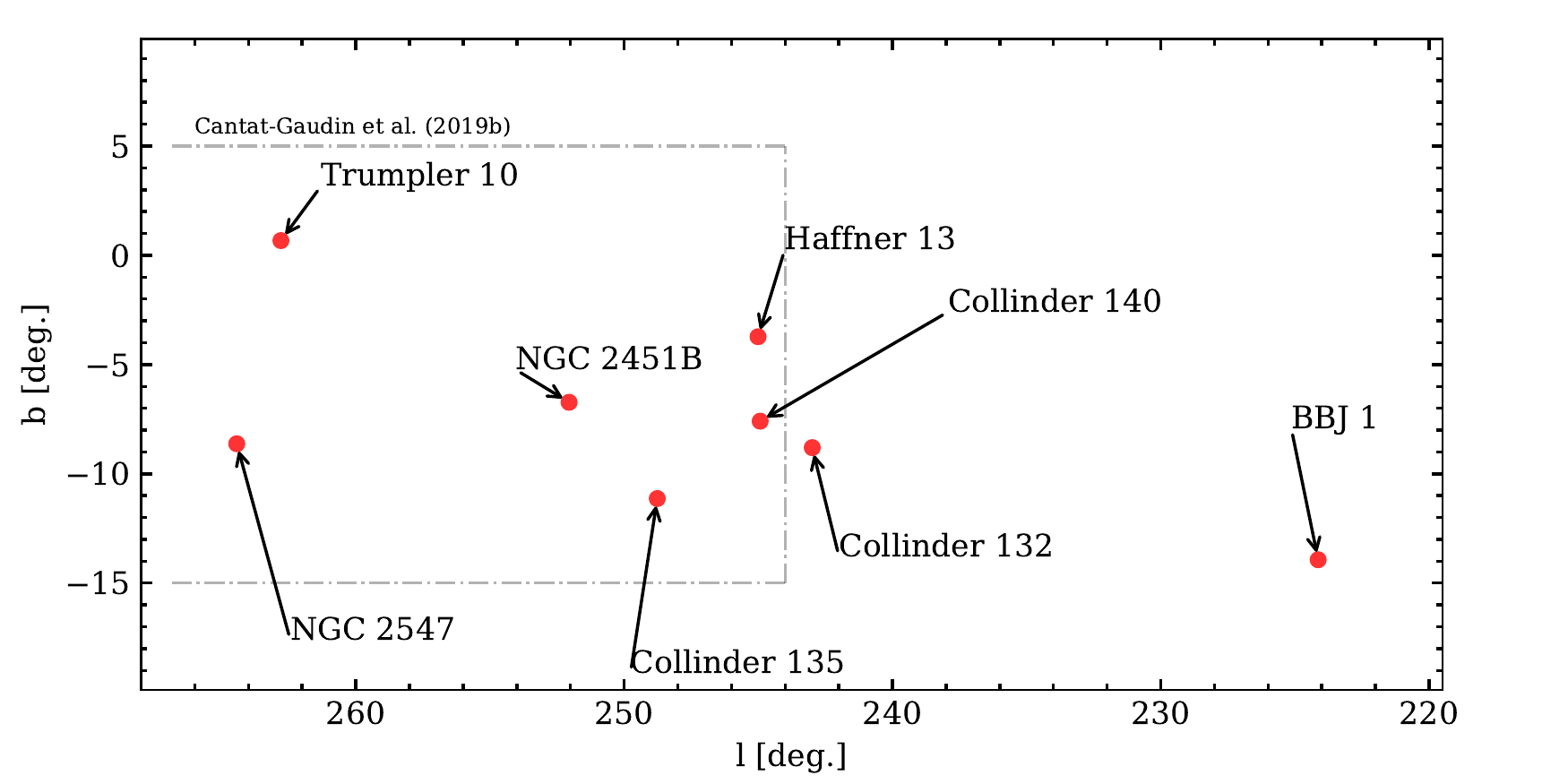}
  \caption{Galactic spatial distribution of the clusters in the region. The dash-dot line indicates the region studied by~\citet[][]{ca19b}.}
  \label{fig:map}
\end{figure*}

As described in the Introduction, the Vela OB2 complex has been the subject of several detailed investigations. In particular
\citet[][]{ca19b} found a complex set of stellar populations whose spatial distribution across 180 pc traces the IRAS Vela Shell and shows clear signs of expansion. The newly discovered cluster is located at a distance of $\sim$259~pc from NGC 2547 and $\sim$158~pc from Collinder 135 (see Fig.~\ref{fig:map}). In the figure we show as a dash-dotted line the region studied by~\citet[][]{ca19b}. As shown in~\citet[][]{gro18}, these two clusters, together with Collinder~140, NGC~2541B and UBC~7, belong to a $\sim$30~Myr stellar complex named Population IV which includes 5 known clusters with common distribution in proper motions and distributed over an are of almost 150 pc on sky. The same authors note that the whole system of clusters appear to be bridged by a diffuse and elongated population of stars that physically connects the mentioned clusters. 

\subsection{Clustering: Step 1}

Admittedly,  BBJ 1 is located at lower Galactic longitude $l$ with respect to the Vela complex as shown in Fig.~\ref{fig:map}. Nonetheless, driven by the intriguing observational fact that the stars in  BBJ 1 and NGC 2547 are coeval, we decided to explore the exquisite astrometric measurements provided by the Gaia DR2 to detect the presence of a filamentary structure of stars possibly bridging BBJ 1 to the whole complex of clusters in the Vela OB2 region.   
We decided to use again the DBSCAN clustering algorithm. We divided the search in three steps. We first looked for star clusters in the area included in the Galactic coordinates $210<l<280$ and $-30<b<10$, where all the targeted clusters are located (see Fig.~\ref{fig:map}). We also restrict the clustering research to the stars in the range of proper motions $-14<\mu_\alpha<-3$ and $2<\mu_\delta<11$ mas/yr, while keeping the original parallaxes range $2<\varpi<3.5$. We run the DBSCAN clustering algorithm using eps=0.135 and min\_samples=30. 
\begin{figure*}
\centering
 \includegraphics[width=1\hsize]{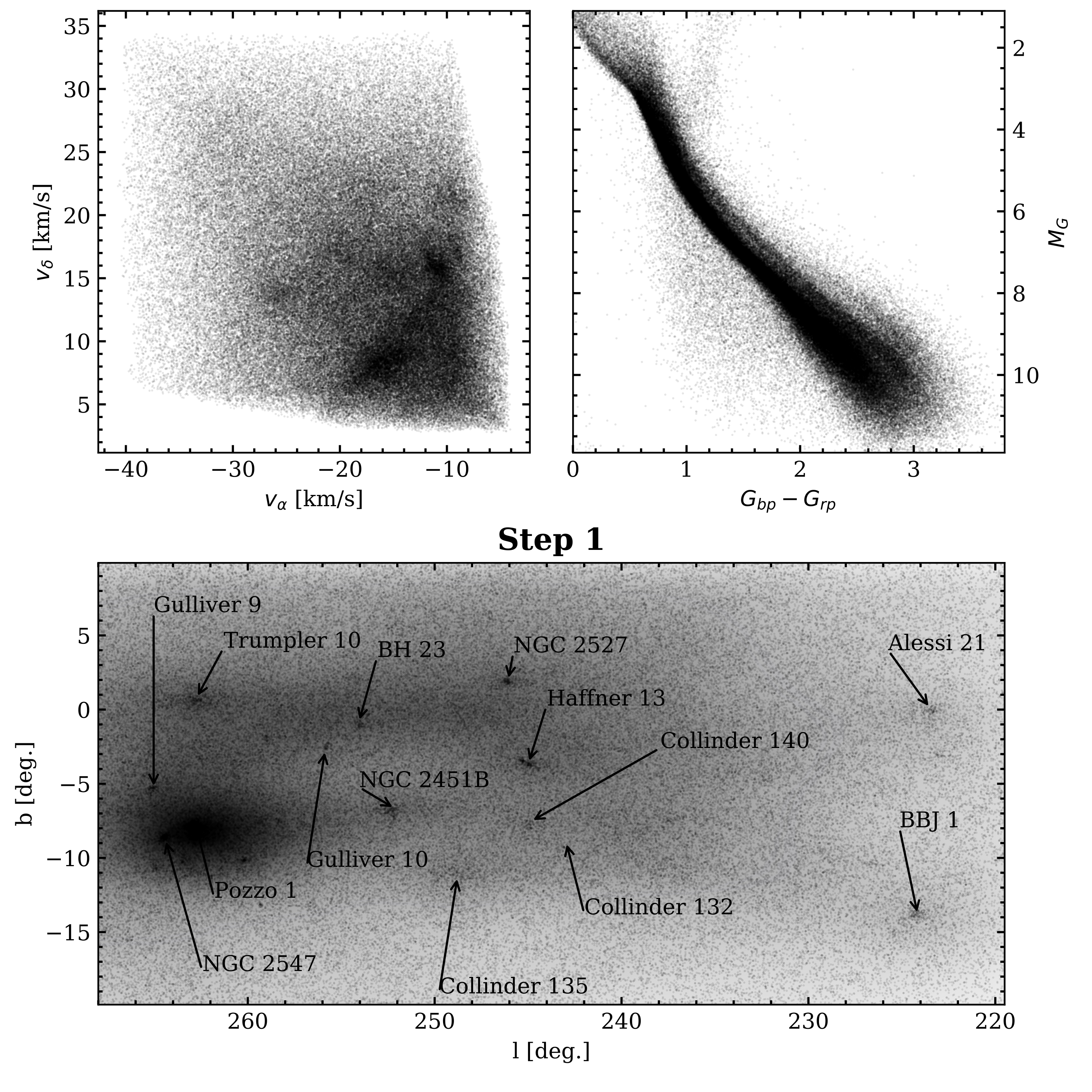}
  \caption{Upper-left: Distribution in the $v_\alpha$ and $v_\delta$ space of the stars in the region after the first iteration with the clustering algorithm. Note that the $\mu_\alpha$ and $\mu_\delta$ are converted from mas/yr into Km/s adopting the individual parallaxes of each star. Upper-right: CMDs of the same stars shown in the top-left panel. Lower-panel: Galactic spatial distribution of the same stars shown in the top panels. The position of the known clusters is also shown.}
  \label{fig:step1}
\end{figure*}

In the lower panel of Fig.~\ref{fig:step1} we show the distribution on sky of the stars isolated by the initial clustering search (black points). We indicated on the map the location of the clusters found by~\citet[][]{cg18} in the same parallax, $v_\alpha$ and $v_\delta$ ranges. Please note that $v_\alpha$ and $v_\delta$ are expressed in Km/s and were estimated as $v_\alpha=\mu_\alpha\times4.74/$parallax and $v_\delta=\mu_\delta\times4.74/$parallax. The two upper panels show the distribution of the proper motions and the CMD of the same stars shown in the map. 

The first step allows us to fully recover all the clusters already known in the region, including the new cluster BBJ 1. The distribution of the proper motions clearly show a set of clusters that are visible in the region.  Intriguingly, while the population of field stars is still strongly present in the catalogue as indicated by the diffuse distribution of stars in the proper motion and sky space, the CMD show the presence of a population stars at age similar to NGC~2457 (and the newly discovered BBJ~1) and possibly a younger component.  An old population of Red Giants is also visible in the plot, certainly composed of stars belonging to the field.

Moreover, a close inspection of the sky map suggests the presence of an elongated structure connecting the stellar population in BBJ 1 to the location of Collinder 140 and Collinder 135. On the same vein, a population of stars seems to bridge the population of the cluster NGC 2547 to Collinder 140. 

\subsection{Clustering: Step 2}

In order to investigate in depth the nature of the distribution of the stars in the map and in the CMD, we perform a second clustering search with DBSCAN on the catalogue obtained after the first step. The main goal of the second step is to remove most of the contamination from field stars that are clearly still present in the catalogue of stars obtained after step 1. For this scope, in this second run we constrain the clustering search in a 6-dimensional box, composed of the $v_\alpha$ and $v_\delta$, the Galactic coordinates $l$ and $b$ and  in the $G_{bp}-G_{rp}$ vs $M_G$ space (i.e. the CMD). This way we force the clustering algorithm to search for stars that are distributed as coeval stellar populations on the CMD while showing common dynamical properties. We run this second session of DBSCAN using the parameters eps=0.6 and min\_samples=30. 


\begin{figure*}
\centering
 \includegraphics[width=1\hsize]{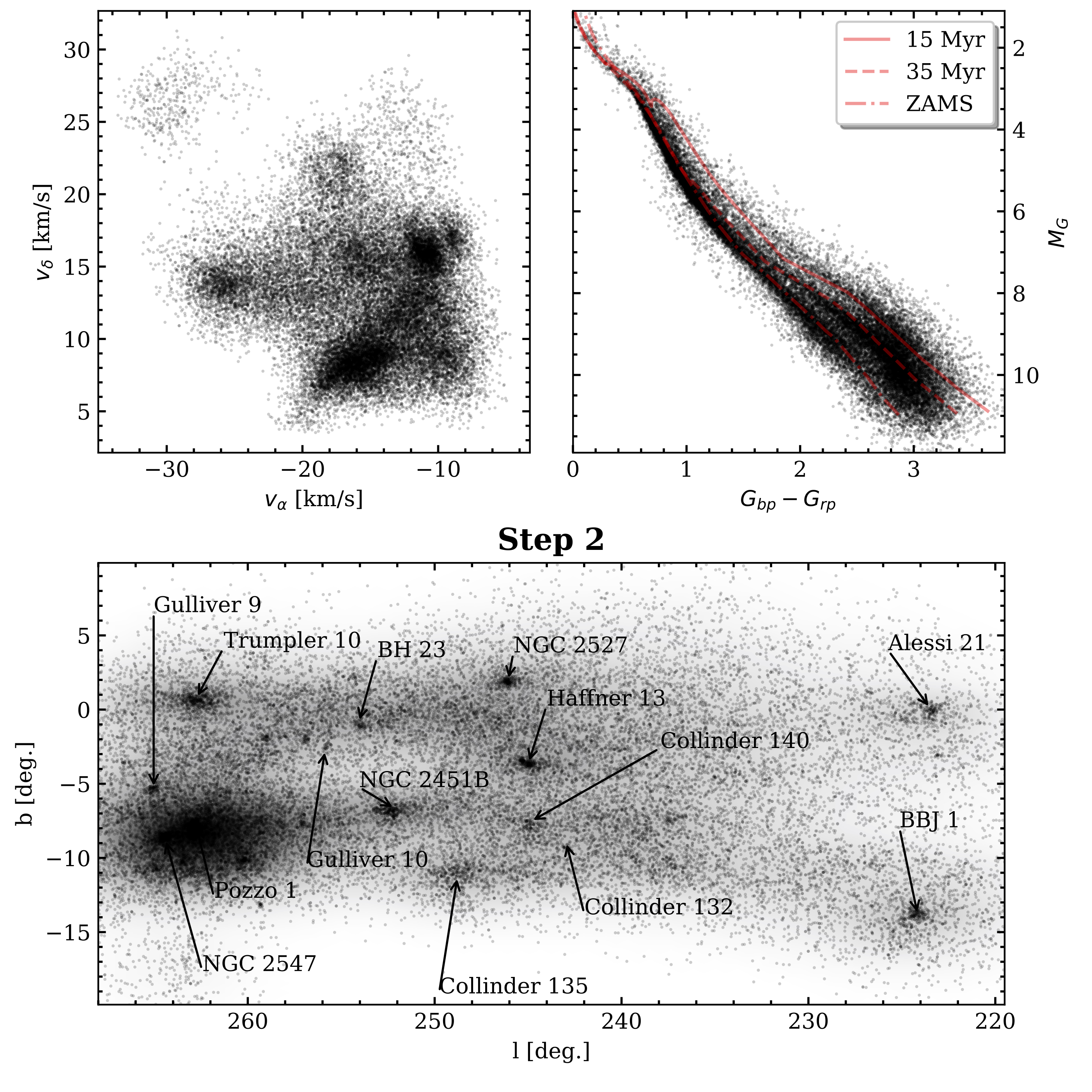}
  \caption{Upper-left: Distribution in the $v_\alpha$ and $v_\delta$ space of the stars in the region after the second iteration with the clustering algorithm. Note that the PM are converted from mas/yr into Km/s adopting the parallax. Upper-right: CMDs of the same stars shown in the top-left panel. Lower-panel: Galactic spatial distribution of the same stars shown in the top panels. The position of the known clusters is also shown.}
  \label{fig:step2}
\end{figure*}

We show in Fig.~\ref{fig:step2} the distribution in $v_\alpha$ and $v_\delta$ (top-left), on the CMD (top-right) and in the space $l$ and $b$ (lower panel) of the stars identified at the second filtering using the DBSCAN clustering. The grey-scale shading that is shown hereafter on the figures is obtained using a density map with a Gaussian kernel. We used the function gaussian\_kde included in SciPy library~\citep[][]{scipy} to calculate the density map. 

The result is quite impressive. While several sub-groups are starting to be quite well isolated in the proper motion space, the distribution of the stars in the CMD is mostly described with only 2 isochrones of 15 Myr and 35 Myr each (dashed and solid lines, respectively) together with the Zero Age Main Sequence (ZAMS) which identifies the oldest population still present in the catalogue. The models shown in the figure are taken from~\citet[][]{bres12} at solar metallicity. We adopted a single extinction of A$_V$=0.124, typical of NGC 2547~\citep[][]{bossini19}. We stress here that, while a dedicated study of the differential extinction might further constraint the possible presence of intrinsic age spreads in the studied population, the adopted isochones provide an excellent service to the present scope of showing our capability to isolate distinct stellar populations. As visible in the sky map, a filament-like structure of stars connecting the several clusters identified in the region is also emerging.  

\subsection{Clustering: Step 3}

While the second step of clustering certainly helped in isolating the clusters in the regions, the CMD of Fig.~\ref{fig:step2} clearly indicates that our selected catalogue is still contaminated by an old component ($>50$ Myr). Such population, well identified via the ZAMS model, is mostly populated by genuine MS field stars belonging to the Galaxy together with the oldest clusters in the region, i.e. Alessi 21 and NGC~2527~\citep[$\sim$70~Myr and $\sim$830~Myr, respectively; see][]{bossini19}. 

We thus run one last clustering filtering (step 3) in the proper motions and $l$ and $b$ spaces with the aim to further separate the stellar populations and isolate the stellar components visible in the CMD shown in Fig.~\ref{fig:step2}. As in step 1, we use eps=0.135 and min\_samples=30.  
\begin{figure*}
\centering
 \includegraphics[width=1\hsize]{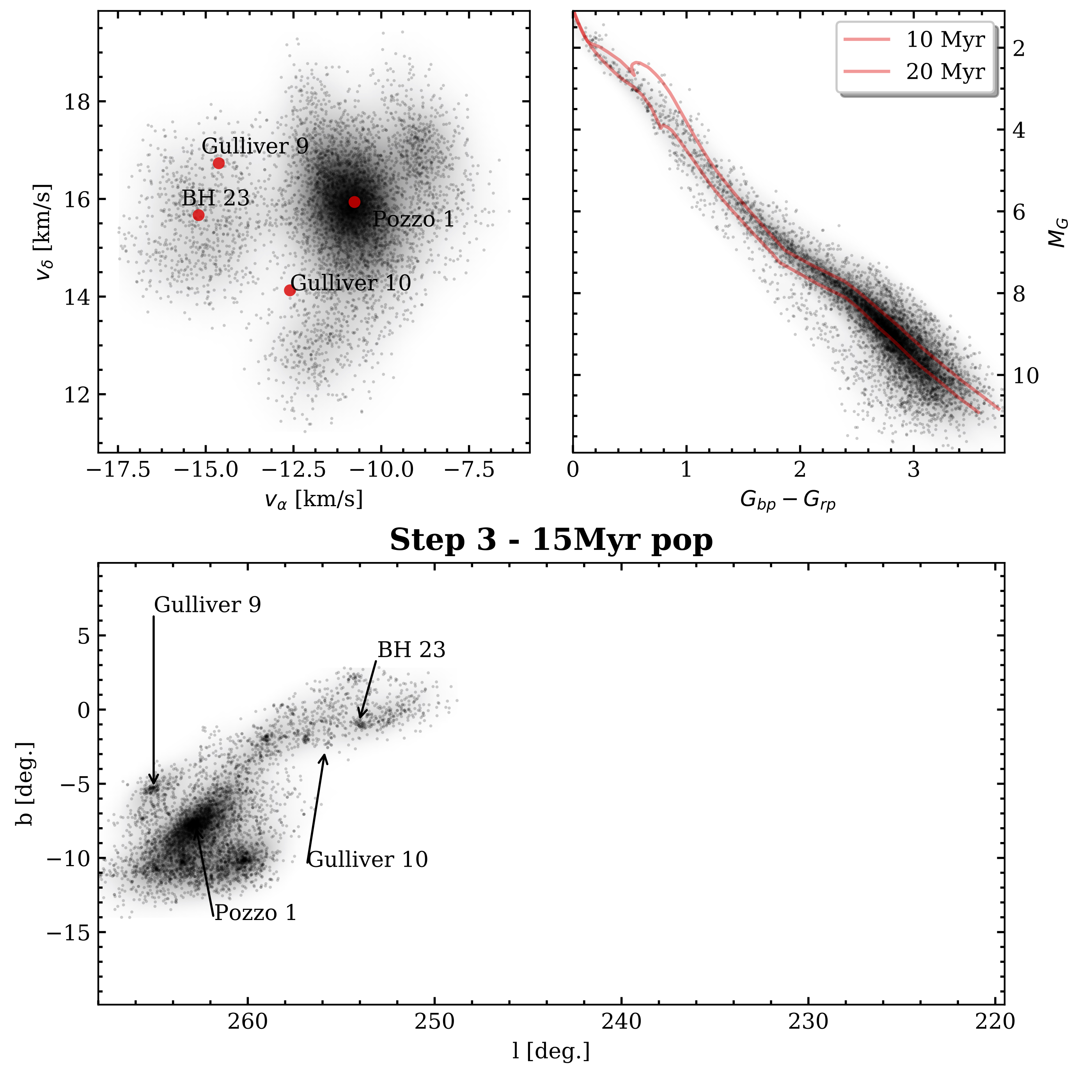}
  \caption{Upper-left: Distribution of the 2D velocities of $\sim15$Myr old stellar component identified in the region after the second iteration with the clustering algorithm. Upper-right: CMDs of the same stars shown in the top-left panel. Lower-panel: Galactic spatial distribution of the same stars shown in the top panels.}
  \label{fig:step3y}
\end{figure*}
\begin{figure*}
\centering
 \includegraphics[width=1\hsize]{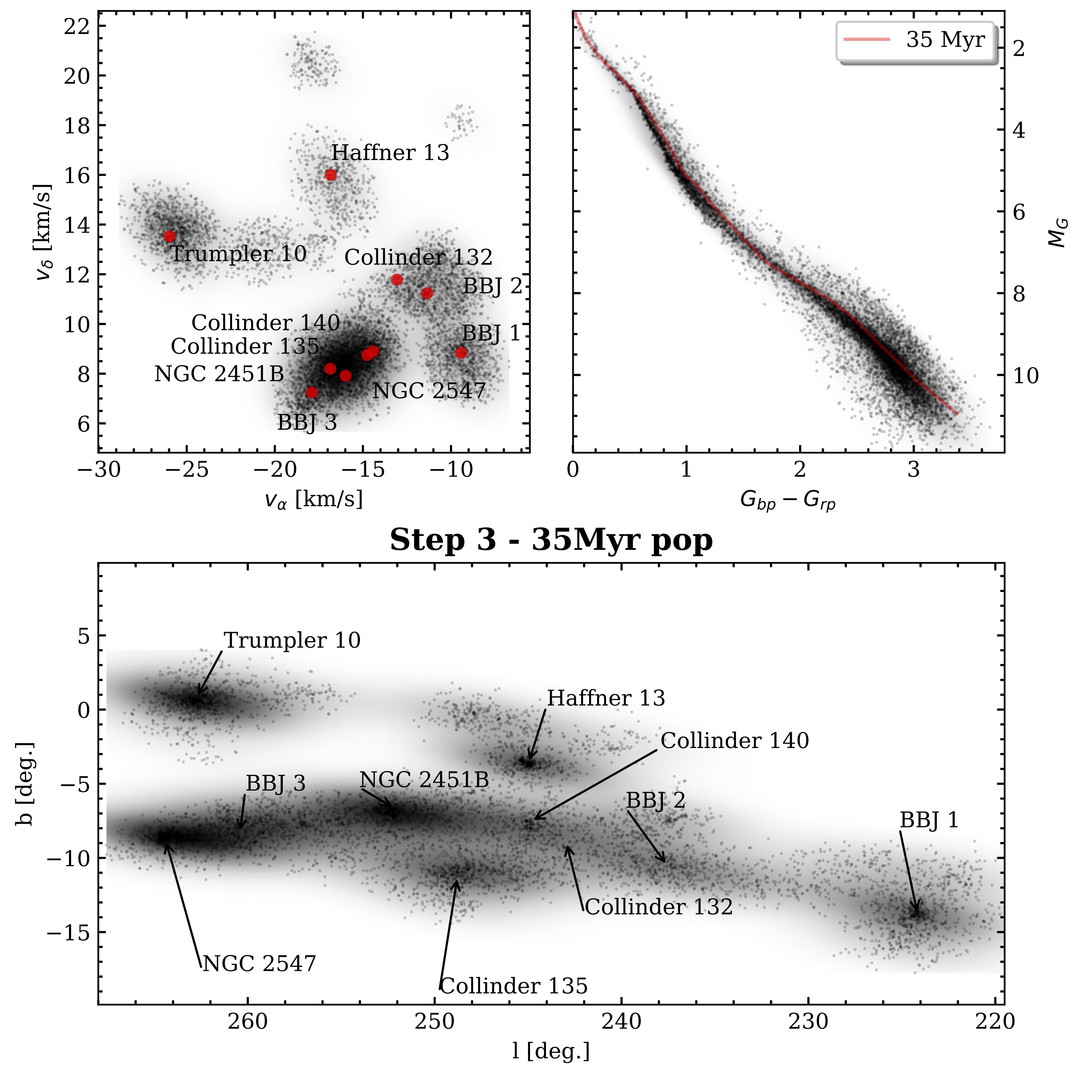}
  \caption{Upper-left: Distribution of the 2D velocities of $\sim35$Myr old stellar component identified in the region after the second iteration with the clustering algorithm. Upper-right: CMDs of the same stars shown in the top-left panel. Lower-panel: Galactic spatial distribution of the same stars shown in the top panels.}
  \label{fig:step3o}
\end{figure*}

We show in Fig.~\ref{fig:step3y} and ~\ref{fig:step3o} the result of the last clustering process. In the two figures we separate the stars according to the age of the population. We stress here that no artificial cut or selection is applied but the populations are shown as they come out of the clustering filtering. We used the re-normalised unit weight error (RUWE) parameter to perform a sanity check of the accuracy of the overall astrometric solution of the stars in the final catalogue (i.e. after the Step 3). The RUWE is expected to be around 1.0 for sources where the single-star model provides a good fit to the astrometric observations\footnote{Please see section 14.1.2 of the GAIA DR2 documentation}. A value significantly greater than 1.0 (say, >1.4) could indicate that the source is non-single or otherwise problematic for the astrometric solution.  We find that 95\% and 93\% of the stars in our final catalogue have RUWE<1.4 and <1.2 respectively, independently of their age. This fact proves the solidity of the astrometric solutions of the stars that populate the final catalogue and that are used hereafter for the analysis.

\begin{figure*}
\centering
 \includegraphics[width=0.42\hsize]{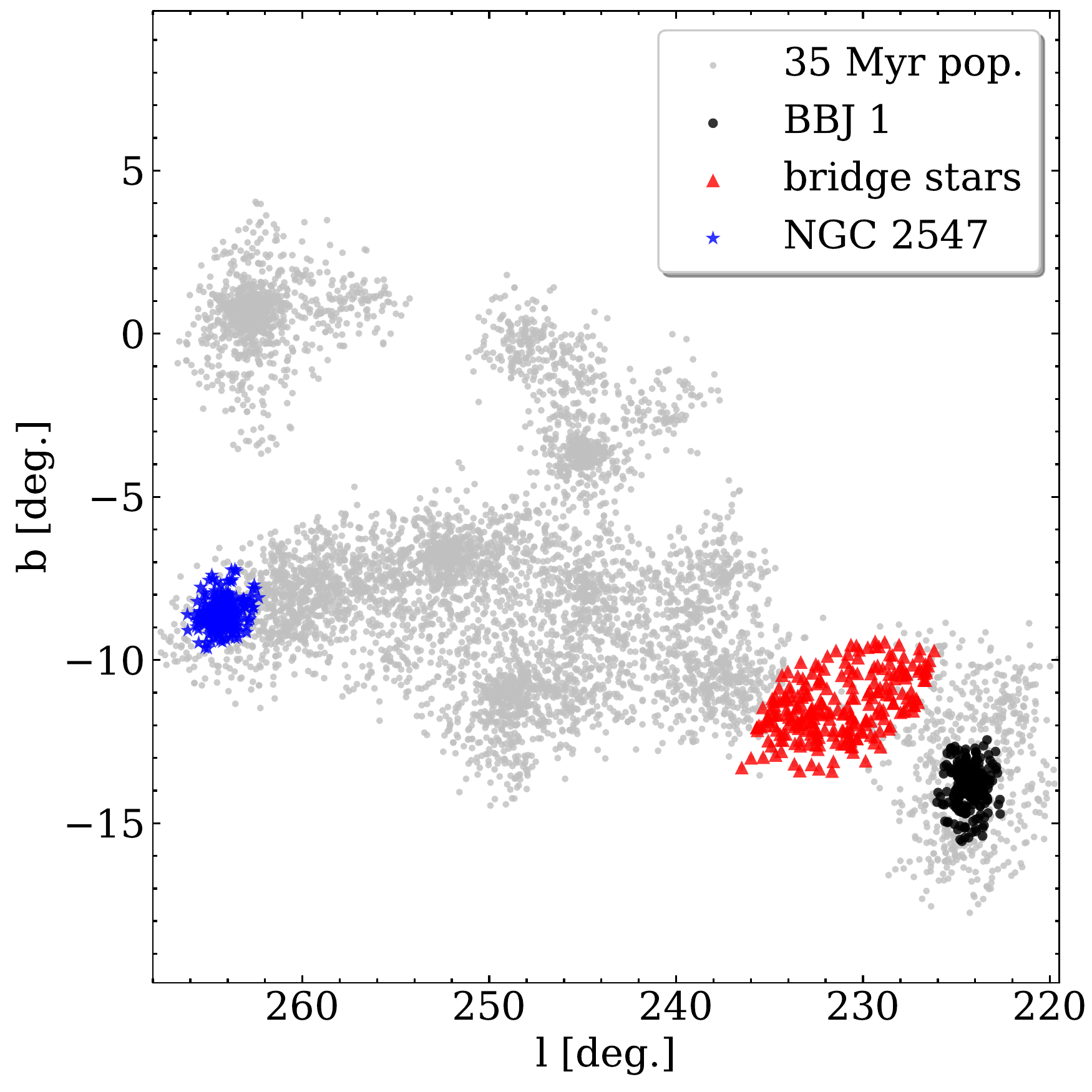}
 \includegraphics[width=0.42\hsize]{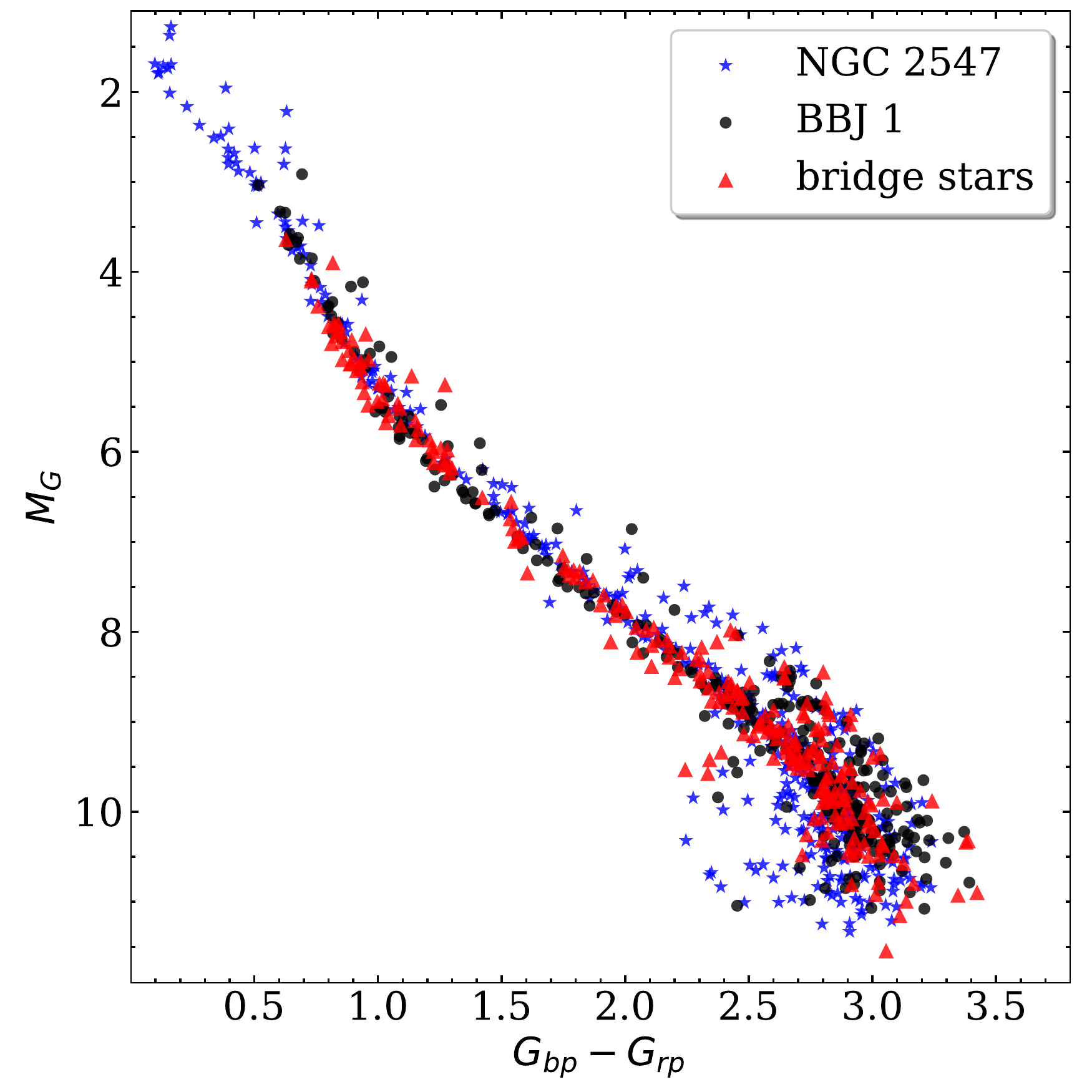}
  \caption{\textit{Left-panel}: Map of the 35 Myr old population isolated with Step 3 selection process (grey points). The blue, black and red points show the stars belonging to NGC~2547, BBJ~1 and a selection of stars along the stellar filament bridging the entire structure.\textit{Right-panel}: Comparison on the CMD of the stars belonging to BBJ 1 (black dots) and NGC 2547 (blue star) and a sample of stars along the stellar filament (red triangle). The stars are clearly coeval. 
  }
  \label{fig:bridge}
\end{figure*}

Through step 3 we perform a remarkable work in isolating coeval stellar populations in the region, a young component of $\sim$15~Myr and an older component of $\sim$35~Myr. Moreover, as shown in the sky map of Fig.~\ref{fig:step3o} (lower panel) we find that a filament of stars of the age of $\sim$35~Myr clearly bridges BBJ 1 at $l\sim$225 to the coeval open cluster NGC~2547 at $l\sim$265. We show in Fig.~\ref{fig:bridge} the position on the sky (left panel) and on the CMD (right panel) of stars belonging to NGC~2547 (blue stars), BBJ~1 (black dots) and a selection of stars along the discovered stellar filament (red triangles). We emphasise here that, as a sanity check, for the latter group we considered only stars with a measured parameter $0.9<$RUWE$<1.1$. The CMD clearly demonstrate that the stars belonging to the three sub-groups are coeval. We stress here that we do not perform a detailed study of the impact of differential reddening to the observed CMD. Our choice is justified by the fact that, as shown in Figure D.1 of ~\citet[][]{ca19b}, in the distance range 0pc to 350pc (where most of our clusters reside) the differential extinction is quite low and mostly affecting the region around Gamma Vel where we find the youngest population that is likely partially embedded in the molecular clouds. Given the extent in sky of the ~35Myr old filamentary structure, the distance of most of the discovered stellar populations from the GammaVel region combined with the remarkable homogeneity (in term of stellar sequences) seen on the CMD of Fig.~\ref{fig:step3o} and~\ref{fig:bridge}, we are convinced that the differential reddening is not severely affecting the magnitudes and colours of the studied populations.

As already shown above, several clusters are found to belong to such elongated stricture. Our clustering algorithm identifies 2 further clusters that we label in the figure as BBJ 2 and BBJ 3. The study of such populations and their nature goes beyond the scope of this paper. Still we mention them as signature of the presence of clumpy sub-structures along the filamentary structure.   

The whole formation is  confined in a range of galactic latitudes between $-17<b<-5$. Remarkably, the young (10 to 15 Myr) stellar component is confined in 2D space in a region $257<l<270$ and $-15<b<-5$ and overlaps with the location on sky of NGC~2547. Such population corresponds to Population VII in~\citet[][]{ca19b}. Finally, the clustering algorithm isolates two very low densities structures in the vicinity of BH~23 (see lower plot) while the old cluster Trumpler 10 seems (according to the distribution in $v_\alpha$ and $v_\delta$) not to be kinematically correlated with the old stellar component previously discussed.       

The figure further confirm the presence of a $\sim$260~pc long filaments of stars with ages between $\sim$30 to $\sim$35~Myr. Part of this population corresponds to Population IV in~\citet[][]{ca19b}. 
It is only thanks to the extension on sky of the current study and the discovery of BBJ~1 that the structure can be seen in its entire extension. It is important to note that we also did the same study with a much larger region to ensure that the coeval populations we found here do not extend further. We can thus confirm that the extend we report here is the maximum one.

%

\section{Relic stellar filament}

As mentioned previously, the Vela OB2 region has been recently the subject of several works aimed at studying the complexity of the stellar populations in the region. In~\citet[][]{becc18} we demonstrated that NGC~2457 and the cluster around the Wolf-Rayet Star $\gamma^2$ Vel belong to a complex constellation of sub-populations confined in a region of $\sim$100~pc$^3$ of volume. Later, \citet[][]{ca19b} extended the study on a wider area of sky in a range of distances from $\sim$250 to $\sim$550~pc. The areas studied include the known clusters Trumpler 10, NGC 2451B, NGC 2547, Collinder 135, Collinder 140 as well as the extended Vela OB2 complex. Using the Gaia DR2 data and detailed clustering analysis they could identify several spatial and kinematic subgroups. Such sub-groups could be separated into seven main groups of ages between 10 and 50 Myr.

Guided by our discovery of a new cluster, BBJ~1, here we expand the study of~\citet[][]{ca19a} and~\citet[][]{ca19b} to a wider area, reaching $l~\sim220$. We find that BBJ~1 is coeval to NGC~2547 but it is located at a distance of $\sim$260~pc from it. We identify a bridge of stars coeval to the two clusters and forming a filamentary structure. Such population was observed already by~\citet[][]{ca19b} as Population IV. {\it In our work this population is revealed to have an extended filamentary structure spanning almost 260pc in length}.

\begin{figure*}
\centering
 \includegraphics[width=1\hsize]{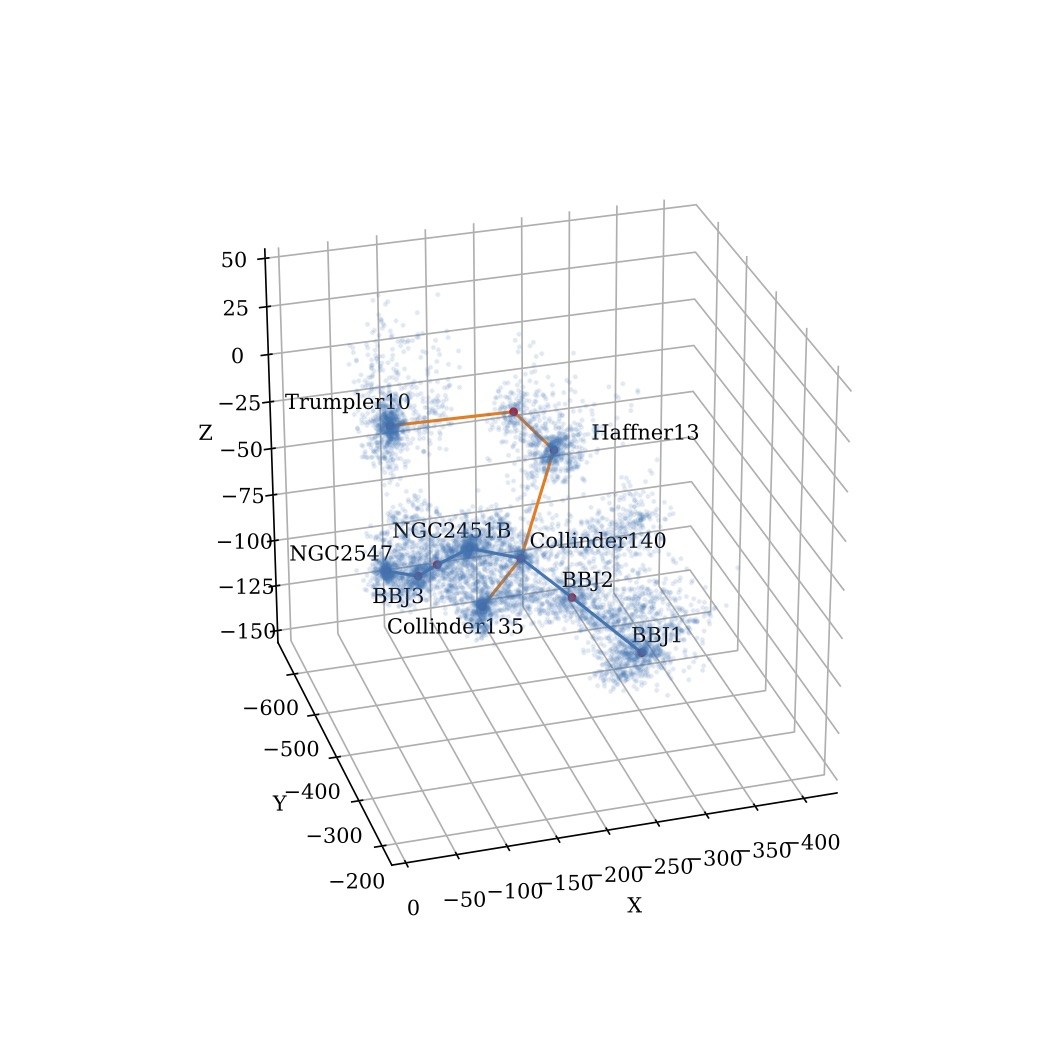}
  \caption{3D distribution of the $\sim$35~Myr population found in this work.}
  \label{fig:3Dold}
\end{figure*}

As discussed in the Introduction, evidence is growing that at the early stages of their formation, pre-stellar cores form in filamentary structures and clusters seem to appear at the intersection of dense filamentary regions in gas holds~\citep[e.g.][]{pl18}. One question would then be how such structures would appear on sky several tens of Myr after the formation of the stars and once the MC is dissipated.
We show in Fig.~\ref{fig:3Dold} the extent and distribution of the old stellar component in the 3D space. A blue line connects the major filamentary structure that is confined in range of distance $\sim$320 to $\sim$420~pc from us while the distance between the clusters are the two extremes of the structures (BBJ 1 and NGC 2547) is $\sim$260~pc. 
~\citet[][]{ca19a} and \citet[][]{ca19b} used the information available with the Gaia DR2 data to study the kinematic structure of the seven populations that were the focus of their study. They find signs of expansion along the Galactic X and Z axes for the whole structure. Still, Population IV that is part of our 35 Myr population show a complex kinematics arrangement that likely reflect the turbulent configuration of the primordial molecular cloud it formed from. 

\begin{figure*}
\centering
 \includegraphics[width=1\hsize]{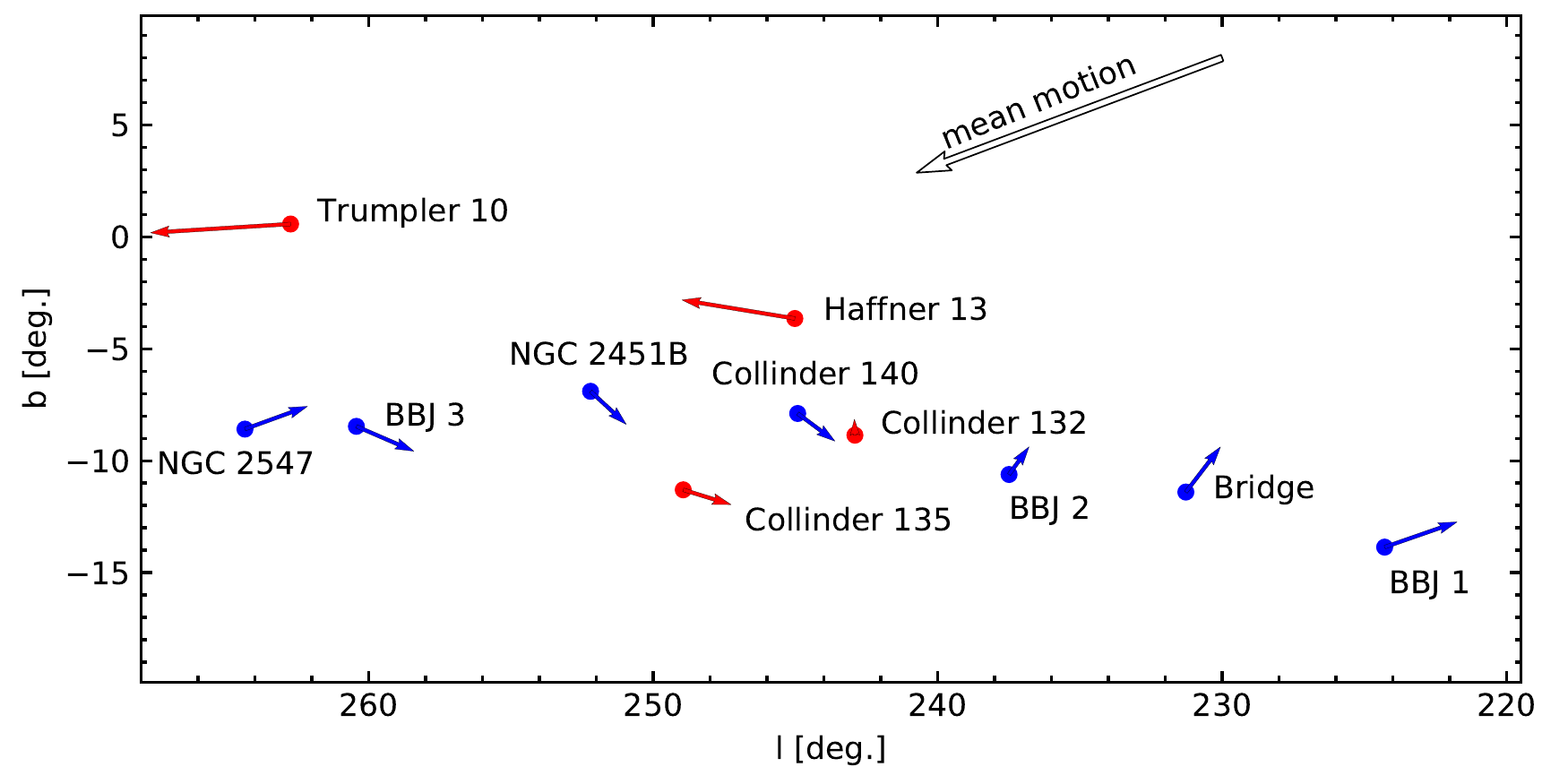}
  \caption{The tangential velocities of the clusters in the region are shown as blue and red arrows. The arrows in the figure are scaled homogeneously. The mean velocity $\mu_{l*}=-15.5\,$km/s, $\mu_b=-5.8\,$km/s (white arrow) of the 35 Myr structure has been subtracted to the motion of the clusters.  In blue we highlights the position and velocity vectors of the clusters along the $\sim35$Myr filament as highlighted in Fig.~\ref{fig:3Dold}.
 }
  \label{fig:vec}
\end{figure*}

While such a scenario requires detailed numerical simulations to be fully understood, we here would like to extend this hypothesis in light of the newly discovered extension of the stellar structure shown in Fig.~\ref{fig:step3o} and~\ref{fig:3Dold}. Using the values of proper motions available with the Gaia DR2 catalogue, we calculate the average tangential velocity of the stellar clusters identified in the $\sim35$~Myr filaments including the stars in the region next to BBJ1 along the filament (named Bridge in Fig.~\ref{fig:bridge}). The mean proper motion of the whole stellar population is $\mu_l=-15.5$km/s, $\mu_b=-5.8 $km/s (white arrow) and has been removed. The result is shown in Fig.~\ref{fig:vec}. We highlight in blue the clusters along the filamentary structure as highlighted in the 3D plot shown in Fig.~\ref{fig:3Dold}. The figure suggests that BBJ1 and NGC2547 were not closer to each other in the past as they seem to have an almost parallel trajectory. At the same time the whole blue structure seems to expand, a fact that confirms what was already indicated in~\citet[][]{ca19b}. Certainly, it would be important to investigate the 3D motion using detailed radial velocity measurements. Unfortunately, the Gaia DR2 catalogue provides radial velocities  for a few stars in the whole region only. We used Eq.6 from~\citet[][]{deVen06} to take into account the perspective expansion/contraction on the tangential velocities used in Fig.~\ref{fig:vec}. We used the mean radial velocities of the stars in each cluster as the value representative of the system itself. While we report that the overall distribution of velocities does not changes significantly from the one shown in Fig.~\ref{fig:vec}, detailed radial velocities measurements are urgently needed to allow us to perform a full 3D kinematic study of the region.

While with the data-set in hand we cannot investigate in more detail its kinematic structure with respect to what already done by~\citet[][]{ca19a,ca19b}, we suggest that the spatial extension ($\sim$260~pc) of the 35 Myr coeval structure discovered in this paper cannot be explained by the mechanism that creates stellar tidal tails and streams around older clusters~\citep[e.g.][]{cabe18stream,ma19stream}. The formation of such tidally formed structures is in fact regulated by internal cluster dynamics, ruled by two-body relaxation that make stars gradually leave the cluster~\citep[][]{bin1987} and the external influence of the gravitational field of the Galaxy that accelerate their dissolution process~\citep[][]{spit1972}. Both mechanisms would lead to the formation of tidally stripped stars in a time-scale of a few hundred relaxation times. 

The young stellar populations presented in this work show a large-scale (260 pc), 35 Myr coeval structure that is too extended and too young to be of tidal origin. We suggest that such large scale structure bear witness of a 35~Myr old snapshot of star formation that happened in a complex and filamentary molecular clouds, like the one recently observed by ALMA~\citep[e.g.][]{Hacar2018}. We call such structure a \textit{relic stellar filament}.

\section{Conclusions}

In this paper we use Gaia~DR2 data to study the stellar population in a region of 40 degrees of radius around the young cluster Collinder 135. The stars are selected in order to have parallaxes $2<\varpi<3.5$ and $\sigma_{\varpi}/\varpi<10\%$. The region includes the Vela OB2 complex of clusters studied by~\citet[][]{ca19a}~\citep[see also][]{becc18} but extends at larger Galactic longitudes ($l\sim220$). Guided by the catalogue of cluster stellar members recently compiled by~\citet[][]{cg18}, we could visually identify the presence of an over-density of stars roughly centered at RA=06:21:08 Dec=-16:10:00 ($l$=224.221369, $b$=-13.866467) which we could not identify with any previously known stellar cluster or association.

We first used the clustering algorithm DBSCAN in 5D ($\alpha$, $\delta$, $\varpi$, $\mu_\alpha$ and $\mu_\delta$) in order to isolate the new stellar cluster. The algorithm isolates 68 stars that are likely members of the same association based on their 3D spatial distribution and proper motions (see Fig.~\ref{fig:bec1}). We compare the CMD of the new cluster stars with the one obtained with the stars belonging to the young cluster NGC~2547. The two CMDs shown in Fig.~\ref{fig:bec1} are remarkably similar. The two clusters are also located at a similar distance from us, i.e. $\sim380$pc. This fact confirms that the stars identified in 5D space via the clustering algorithm belong to a single stellar population of $\sim$30-35~Myr. We call the new cluster BBJ 1.

The comparison of the CMDs shown in the right panel of Fig.~\ref{fig:bec1} demonstrate that NGC 2547 and BBJ 1 are coeval. In their recent work, ~\citet[][]{ca19a,ca19b} showed that NGC 2547 belongs to a coeval ($\sim$30~Myr) family of clusters which are grouped in space at the center of a expanding IRAS Vela Shell. The subgroup includes the clusters Collinder 135, Collinder 140, NGC 2451B and NGC 2547. The whole population extends in Galactic longitude from $l\sim245$ to $l\sim265$ and latitude $b\sim-15$ to $b\sim-5$~\citep[see figure 1 from][]{ca19b}.

Remarkably,~\citet[][]{je19} using Gaia DR2 data could isolate a previously unknown 17 Myr-old stellar substructure in the Orion star forming region (OSFR). Located at a distance of 430 pc from us, this system has a filament-like shape on the sky of length of $\sim$90~pc. Radial velocity follow-up observations of the stars in the system confirmed that the group of stars share a common kinematic. The authors call such system the Orion relic filament suggesting that, as for the relic filament shown in this work, the young age and extent of the structure suggest an in-situ formation of the stars in a filamentary star formation event.

From a theoretical point of view,~\citet[][]{zuc19} recently showed the presence of large-scale filaments of $>100$pc size in AREPO simulations that can form in Giant Molecular Clouds by invoking simple galactic dynamics. 
Smith et al. (submitted to MNRAS) used a set of new simulations which self-consistently 
forms molecular cloud complexes at high enough resolution to resolve internal substructure all while including galactic-scale forces. 
These new simulations offer new support on the formation of stars in giant molecular clouds along filamentary structure of several tenth of
parsec in length. A detailed study of the kinematical properties of the stellar populations found in our current work (possibly including detailed
radial velocities) will provide valuable observational constraints to allow us to infer the initial condition of the formation of such structures.\\
Very recently~\citet[][]{kou19} used unsupervised machine learning techniques on the Gaia DR2 catalogue to identify stellar clusters, associations, and co-moving groups within 1 kpc and $|b|<30^{\circ}$. In great agreement with our results, they find groups of stars with ages <100Myr with a filamentary or string-like structure. By inspecting their Figure 10 and the interactive figures~\footnote{http://mkounkel.com/mw3d/} we can confirm that the same structure detected in our work is also at least partially detected in their fully independent work. In agreement with our conclusion they also suggest that these strings are primordial and are not formed as a result of a tidal disruption of the clusters. 

To answer the question asked in the introduction:  what imprint does the star formation in seemingly large scale filaments leave in young stellar populations just emerging from their natal molecular clouds? The young stellar population discovered in this work and in Orion seem to suggest the presence of stellar structures that keep a memory of their filamentary star formation. Such assembly of stars are too young to be associated to tidal streams. While a solid conclusion on the origin of such structure requires detailed radial velocity measurements able to reconstruct its 3D kinematics, we propose that it represents the relics stellar structure of star-formation happening in filamentary molecular clouds.

\section*{Acknowledgements}
This work has made use of data from the European Space Agency (ESA) mission
{\it Gaia} (\url{https://www.cosmos.esa.int/gaia}), processed by the {\it Gaia}
Data Processing and Analysis Consortium (DPAC,
\url{https://www.cosmos.esa.int/web/gaia/dpac/consortium}). Funding for the DPAC
has been provided by national institutions, in particular the institutions
participating in the {\it Gaia} Multilateral Agreement.
This research has made use of the SIMBAD database,
operated at CDS, Strasbourg, France.
TJ acknowledges support by the Erasmus+ programme of the European Union under grant number 2017-1-CZ01- KA203-035562.



\bibliographystyle{mnras}
\bibliography{paper} 




\appendix
\section{The clusters kinematics}

\begin{figure}
\centering
 \includegraphics[width=1\hsize]{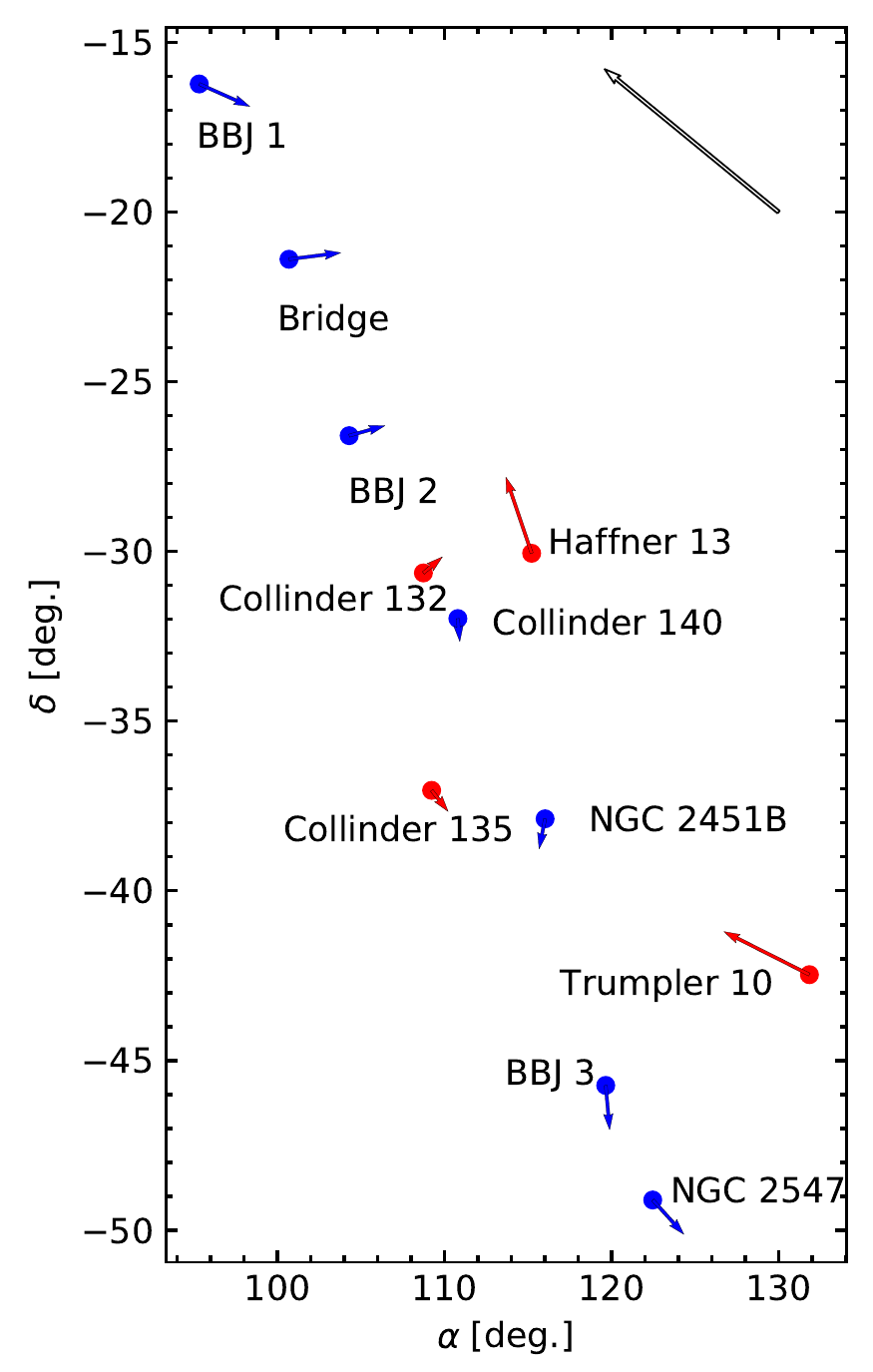}
  \caption{The tangential velocities of the clusters in the region are shown in the right ascension (ra) and declination (dec) plane as blue and red arrows. The arrows in the figure are scaled homogeneously. The mean velocity (white arrow), $\mu_{\alpha*} = -12.8\,$km/s, $\mu_{\delta} = 10.5\,$ km/s, of the 35 Myr structure has been subtracted to the motion of the clusters.  In blue we highlights the position and velocity vectors of the clusters along the $\sim35$Myr filament as highlighted in Fig.~\ref{fig:vec}.
 }
  \label{fig:vecradec}
\end{figure}

\begin{figure}
\centering
 \includegraphics[width=1\hsize]{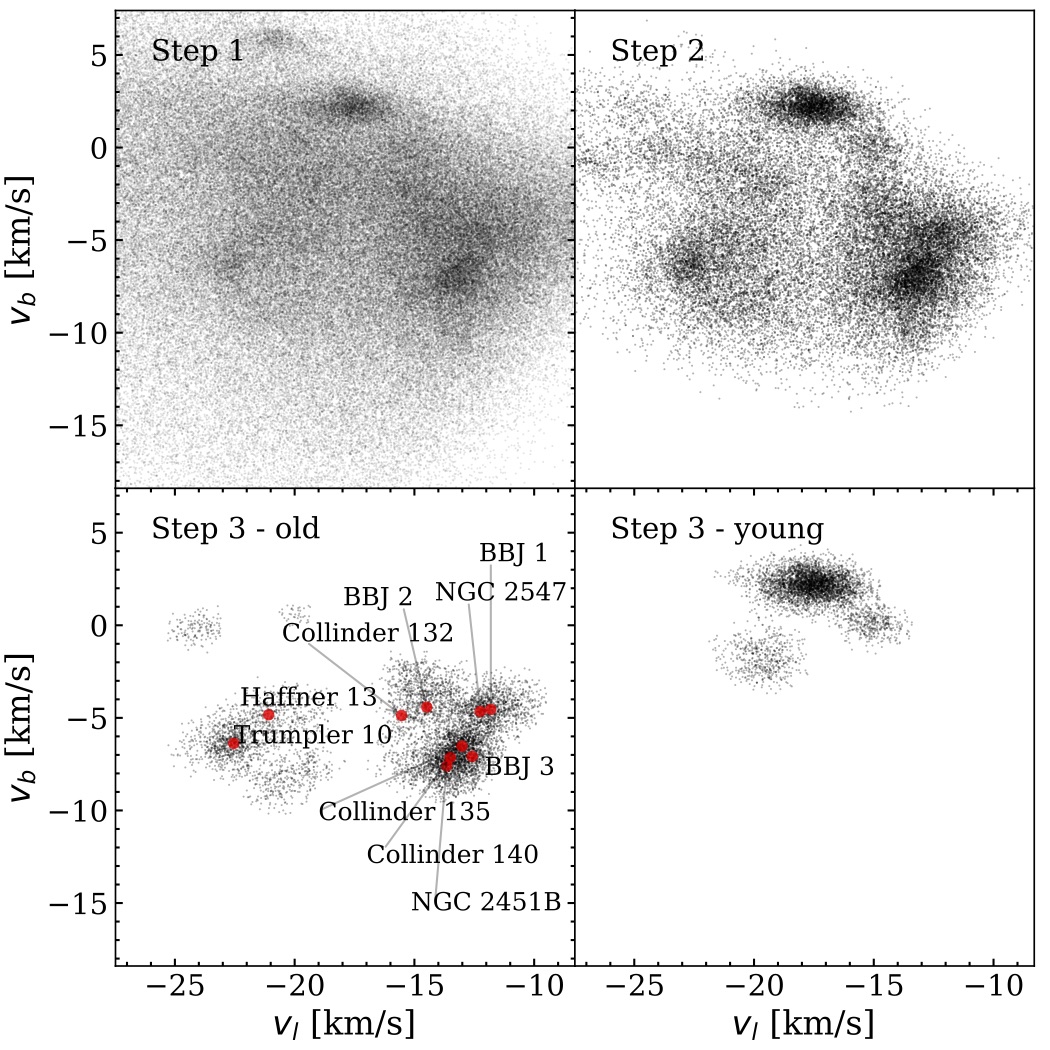}
  \caption{The distribution of the proper motions of the stars in the region in the $l$ and $b$ as they appear after each clustering step as described in Sec.~\ref{sec:clustering}.
 }
  \label{fig:veclb}
\end{figure}

In Fig.~\ref{fig:step1}, \ref{fig:step2}, \ref{fig:step3y} and ~\ref{fig:step3o} of the manuscript we show the distribution of the 2D velocities in km/s in the right ascension ($\alpha$) and declination ($\delta$) plane. This choice is driven by the fact that such plain was effectively used by the DBSCAN clustering algorithm to isolate the stellar populations. Here we show in Fig.~\ref{fig:vecradec} the tangential velocities of the clusters in the same plane to offer a view of the 2D kinematic consistent with the plane used in the selection process. On the same vein, we show in Fig.~\ref{fig:veclb} the distribution of the proper motions of the entire stellar population in the $l$ and $b$ plane as it evolves during the selection steps. These two plots consistently indicate that the difference in velocity of the different populations belonging to the old filamentary structure (blue arrows on Fig.~\ref{fig:vecradec}) is of the order of few km/s. On the contrary Trumpler 10 and Haffner 13 seem to be disconnected from the main structure, in line with their 3D spatial location shown in Fig.~\ref{fig:3Dold}.
In principle, with a differential velocity of few km/s some of the systems might have been much more close to each other at the early stages of their formation (i.e.$\sim35$Myr ago). Still, according to the 2D velocity vectors shown in Fig.~\ref{fig:vecradec} and Fig.~\ref{fig:vec} it seems difficult to trace the cluster back to a limited and well confined spacial location.
As mentioned in the paper, detailed measurements of the radial velocities of the different populations are urgently needed to assess the realistic 3D kinematic of the entire structure.

%


\bsp	
\label{lastpage}
\end{document}